\documentclass[letterpaper]{article}
\usepackage[top=3cm,bottom=3cm,left=3cm,right=3cm,marginparwidth=1.75cm]{geometry}
\usepackage{jheppub}  
\usepackage{listings}
\usepackage{xcolor}
\usepackage{multirow}
\usepackage{natbib}
\usepackage{dcolumn}
\usepackage[english]{babel}
\usepackage[utf8x]{inputenc}
\usepackage[T1]{fontenc}
\usepackage{palatino}
\usepackage{amsmath}
\usepackage{afterpage}
\usepackage{graphicx}
\usepackage[colorinlistoftodos]{todonotes}
\usepackage{makeidx}
\newcommand{\bea}{\begin{eqnarray}}  
\newcommand{\eea}{\end{eqnarray}}  
\makeindex
\begin{document}

\vspace*{1.2cm}

\thispagestyle{empty}
\begin{center}

{\LARGE \bf Most dwarf spheroidal galaxies surrounding the Milky Way cannot be dark-matter dominated satellites }

\par\vspace*{7mm}\par

{

\bigskip

\large \bf Francois Hammer, Yanbin Yang, Frederic Arenou, Hefan Li, Jianling Wang, Piercarlo Bonifacio, Carine Babusiaux, Yongjun Jiao}

\bigskip

{\large \bf  e-mail: francois.hammer@obspm.fr}

\bigskip

{ GEPI, Observatoire de Paris, Universit\'e PSL, CNRS, Place Jules Janssen 92195, Meudon, France; NAOC, Chinese Academy of Sciences, A20 Datun Road, 100012 Beijing, PR China}

\bigskip

{\it Presented at the 3rd World Summit on Exploring the Dark Side of the Universe \\Guadeloupe Islands, March 9-13 2020}

\end{center}

{  \bf  Abstract }

\vspace*{1mm}

\begin{abstract}

Milky Way dwarf spheroidal galaxies are the tiniest observed galaxies and are currently associated with the largest fractions of dark matter, which is revealed by their too large velocity dispersions. However, most of them are  found near their orbital pericenters. This leads to a very low probability, P = 2 $10^{-7}$, that they could be long-lived satellites such as sub-halos predicted by cosmological simulations. Their proximity to their pericenters suggests instead that they are affected by tidal shocks, which provide sufficient kinematic energy to explain their high velocity dispersions. Dependency of the dark matter properties to their distance to the Milky Way appears to favor tidally shocked and out of equilibrium dSphs instead of self-equilibrium systems dominated by dark matter.

\end{abstract}
  
\section{Introduction}
\label{intro}
The Milky Way (MW) halo is populated by many  dwarf spheroidal galaxies (dSphs) that show too large stellar velocity dispersions if considered at  equilibrium within the gravitational attraction of their stellar mass. The pioneering work of ~\cite{Aaronson1983} showed the large velocity dispersion for the Draco stars, which requires significant amounts of DM if the Draco system is considered at equilibrium. Further works have provided robust measurements of velocity dispersion along the line of sight ($\sigma_{\rm los}$) for several tens of MW dSphs \cite{Munoz2018,Fritz2018}, confirming their large amplitude. In the current $\Lambda$CDM scenario, dSphs belong to the population of dark-matter (DM) dominated sub-halos that are very numerous in cosmological simulations. Therefore, the DM content is calculated assuming self-equilibrium  \cite{Walker2009,Wolf2010}, which is also necessary to avoid the destruction of long-lived MW satellites by MW tides. Furthermore, three (UMi, Draco and Sculptor) out of the nine classical dSphs,  contain only old stars, which supports an early accretion into the MW halo.\\
Because of their proximity, MW dSphs are the tiniest galactic objects that can be observed and studied, down to stellar mass as small as few hundred of solar masses. This, associated to their kinematics, let the ultra faint dwarfs (UFDs) to be considered the most DM-dominated objects with DM to baryon mass ratios larger than several hundreds, even up to few thousands. However, \cite{Hammer2018} showed that knowing the gravitational attraction exerted by the MW together with the dSph scale-length ($r_{half}$) and stellar mass, one may accurately derive their DM to baryon mass ratios, which appears at odd with the self-equilibrium scenario. Furthermore, it has been shown that tidal shocks exerted by the MW halo may reproduce the kinematics of dSphs as well. Since DM calculations are based on line of sight measurements, one may wonder whether they could have been corrupted by a physical effect expected if dSphs were out of equilibrium because of MW tides \cite{Hammer2018,Hammer2019,Hammer2020}. 

\section{Consequences of the Gaia 2nd data release (DR2)}
\label{GAIA}
Gaia DR2 has provided accurate proper motions allowing for the first time to:
\begin{itemize}
\item calculate the circular velocity curve of the MW up to 25 kpc \cite{Eilers2019,Mroz2019};
\item derive 3D velocities for most UFDs \cite{Fritz2018,Simon2018},while they were known only for the nine classical dSphs.
\end{itemize}
The first result is based on thorough analysis of a very large sample of 26,000 RGB stars in the MW disk \cite{Eilers2019} resulting into a slightly but robustly determined decline of the circular velocity from 5 to 25 kpc. Such a result is confirmed by \cite{Mroz2019} using 773 Classical Cepheids with precise distances. Subsequent analyses of these, including accounting for the different types of errors, have lead to MW total mass near or well below $10^{12}$$M_{\odot}$ \cite{Eilers2019,deSalas2019,Grand2019,Karukes2020}. This excludes former models with very large MW mass (e.g., from \cite{McMillan2017} or the "heavy" MW model of \cite{Fritz2018}), i.e., those which were able to keep bound Leo I dSph or the Large Magellanic Cloud (LMC).  

The second result have revealed that a large majority of dSphs have eccentric orbits with eccentricities in excess of 0.66 \cite{Fritz2018,Hammer2019} for two third of them. Even if most dSphs appear to be bound to the MW \cite{Fritz2018}, their 3D velocities are very large and consistent with a recent infall, less than 4 Gyr ago (see Figure 2 of \cite{Hammer2020}).

Combined together, the two results lead us to further investigate the nature of dSphs on the basis of their orbital motions.

\section{Most dSphs lie near their pericenters}
\label{peri}
\begin{figure}[ht!]
\begin{center}
\includegraphics[width=0.4\textwidth]{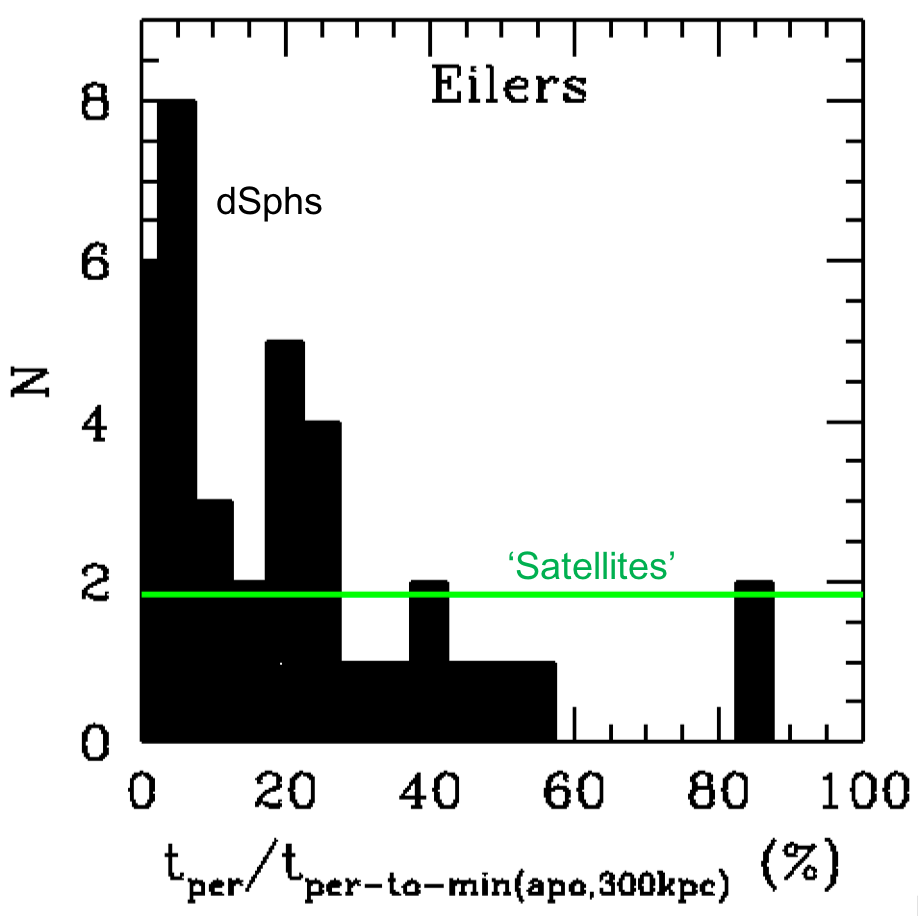}
\includegraphics[width=0.4\textwidth]{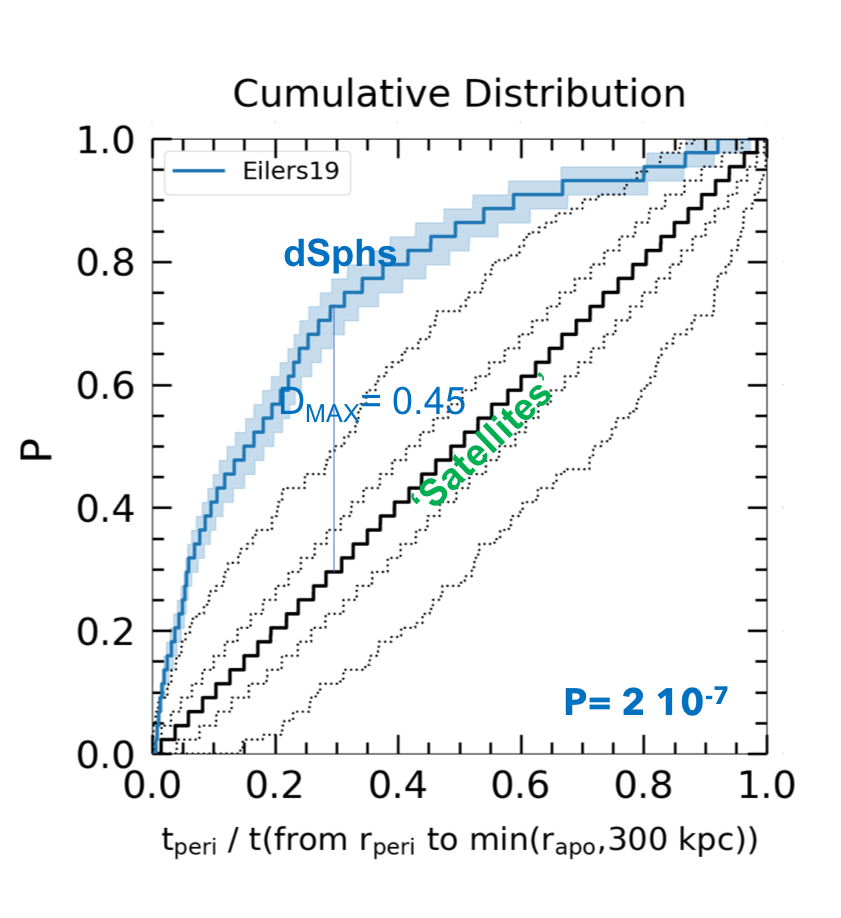}
\end{center}
\caption{\label{fig1} {\it Left:} histogram of the time ratio distribution of dSphs compared to the expectations if they were MW satellites (horizontal green line). {\it Right:} Cumulative distribution of time ratios (blue line and area  showing the 1$\sigma$ error) compared to the null hypothesis, i.e., a randomization of locations expected if dSphs are MW satellites (black solid line).   }
\end{figure}

Pericenter is one of the orbital quantity that can be robustly determined and that slightly depends on the MW mass profile \cite{Simon2019}.  Its determination becomes even more precise if heavy MW mass models are excluded.  Both \cite{Fritz2018,Simon2018} noticed that many dSphs lie near their pericenters. This has been further investigated by \cite{Hammer2020} on the basis of the MW mass model from \cite{Bovy2015} and using the sample from \cite{Fritz2018} (see details in \cite{Hammer2020}). Figure~\ref{fig1} presents results from the same calculations but using instead the MW mass model from \cite{Eilers2019} that is very similar than that of \cite{Bovy2015}, with a difference of less than 10\% for the cumulative mass at all radii.

If dSphs were MW satellites, it would be unlikely to find most of them near their pericenters. To illustrate this, one could remind that comets with eccentric orbits are very rarely found near their pericenter, obeying the second Kepler law. For example, the Halley comet has a period of 77 years and is only seen few months near pericenter. In an extended mass potential such as that of the MW, one has to consider an orbital motion along a rosette instead of an ellipse. Here we estimate the probability that dSphs assumed to be MW satellites could be mostly observed near their pericenters, by performing the following steps:
\begin{itemize}
\item We first compare two samples: (1) is made by virtual, randomly selected "satellites" of the MW and, (2) corresponds to the observed dSphs (see \cite{Hammer2020});
\item The time to reach pericenter divided by the total time from pericenter to apocenter provides the chance occurence for a dSph to be at its position within its orbit, which has been calculated using galpy \cite{Bovy2015} and the mass model from \cite{Eilers2019};
\item Since most dSphs (down to UFD mass, e.g., Coma Berenices) can be seen by the {\it Dark Energy Survey} (DES) up to 300 kpc \cite{Simon2019,Drlica-Wagner2020}, we further limit investigations for both samples to a maximal distance equal to min(apocenter, 300 kpc).
\end{itemize}
 The histogram of the time ratio distribution of dSphs (left panel of Figure~\ref{fig1})  shows how they concentrate near their pericenters contrary to expectations if they were long-lived satellites of the MW (see the green horizontal line). Right panel shows the cumulative distribution to which we have applied a Kolmogorov Smirnov test, which leads to a very low probability (P= 2 $10^{-7}$ from $D_{MAX}$=0.45 for 35 objects, see also \cite{Hammer2020}) that dSphs could be satellites orbiting around the MW. In other words a scenario of long-lived satellites for which DM is shielding dSphs from MW tides is very unlikely.

\section{Are dSphs tidally shocked near their pericenters?}
\label{shocks}
The proximity of dSphs to their pericenters suggest that MW tides are at work. However, \cite{Battaglia2013} have excluded that dSphs could be tidally disrupted DM free galaxies because of their observed internal kinematics and structural properties. This conclusion is based on modeling \cite{Piatek1995,Kroupa1997,Read2006,Iorio2019} that accounts for tidal stripping, which is the dominant process when the satellite is only made of stars and DM (see \cite{Hammer2019,Hammer2020}). In particular, most dSphs do not show tidal disturbance such as tails as it would be expected for tidally stripped galaxies.

The above conclusion could be different if most dSphs were affected by tidal shocks instead of tidal stripping. Let us consider the global instantaneous energy change $\Delta E$ caused by MW tidal forces on an individual dSph star with velocity $\bf v$, which is:

\begin{equation}
\label{energy}
\Delta E = {\bf v \cdot \Delta v} + 1/2 (\Delta v)^2. 
\end{equation}

Tidal shocks may dominate  the energy exchange if the first term vanishes because of spherical symmetry. This had been verified by  \cite{Aguilar1985} for globular clusters affected by the galactic bulge or disk.
In such a case, the latter term (called "tidal shocking" or "heating") is approximated to $1/2 (\Delta \sigma_{los}^2)$\footnote{$\Delta \sigma_{los}^2$=$\sigma_{los}^2$-$\sigma_{stars}^2$ \cite{Hammer2019}, where $\sigma_{los}$ and $\sigma_{stars}$  are the measured velocity dispersion and the 1D los velocity dispersion associated to the sole stellar component, respectively.}, i.e., tidal shocks bring an increase of the kinetic energy for, e.g., stars at r=$r_{half}$, which follows \cite{Hammer2020}: 
\begin{equation}
\label{shock}
1/2 \Delta \sigma_{\rm los}^2=\Delta \Phi_{\rm MW}  =  \sqrt{2} \times g_{\rm MW} \times r_{\rm half}  \times f_{\rm MWshocks},
\end{equation}
where $g_{\rm MW}$ is the gravitational acceleration of the MW and $f_{\rm MWshocks}$ represents the fraction of the system that is affected by tidal shocks \cite{Weinberg1994}. The main discoveries of \cite{Hammer2019,Hammer2020} are:
\begin{itemize}
\item That tidal shocks may also apply on DM-free dSph galaxies under the impulse and distant-tide approximations, in particular if dSphs are at first infall; in such a case it is expected that the main initial component, the gas, has been stripped before the pericenter passage, letting residual stars expanding in a spherical geometry due to the subsequent lack of gravity;
\item  That the kinetic energy change by 1/2 $\Delta \sigma_{\rm los}^2$ along the direction to the Galactic center coinciding to that of the line of sight is precisely what has been assumed to calculate the DM content in dSphs by \cite{Walker2009,Wolf2010}.
\end{itemize}

 Calculation of the DM mass is based on a line of sight measurement, and only the projected mass density ($M_{DM}/r_{\rm half}^{2} $) is known from observations of $\sigma_{\rm los}$. When multiplied by the gravitational constant G, it leads to  the gravitational attraction of the DM exerted on a star at r=$r_{half}$, which is:
\begin{equation}
\label{DM}
a_{\rm DM}= G M_{\rm DM} \times r_{\rm half}^{-2} = \Delta \sigma_{\rm los}^2 \times r_{\rm half}^{-1}.
\end{equation}
 
\begin{figure}[ht!]
\begin{center}
\includegraphics[width=0.5\textwidth]{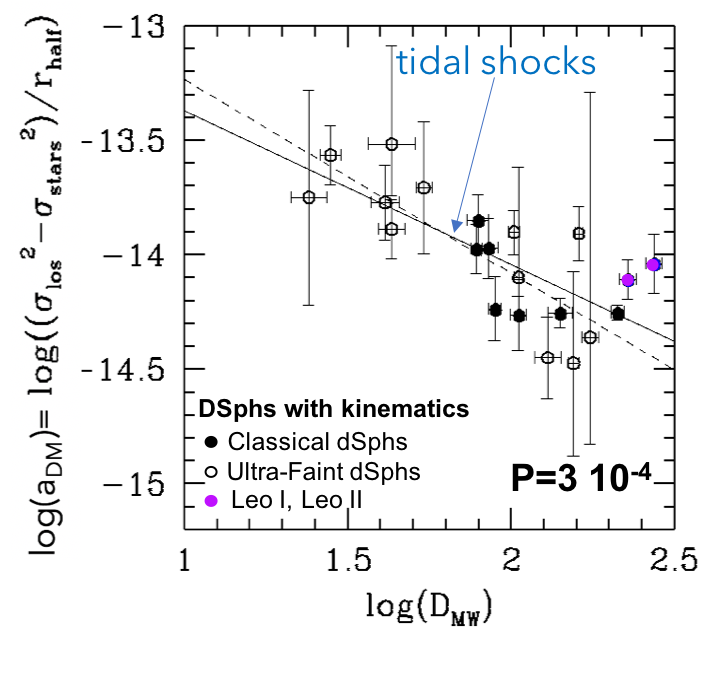}
\includegraphics[width=0.45\textwidth]{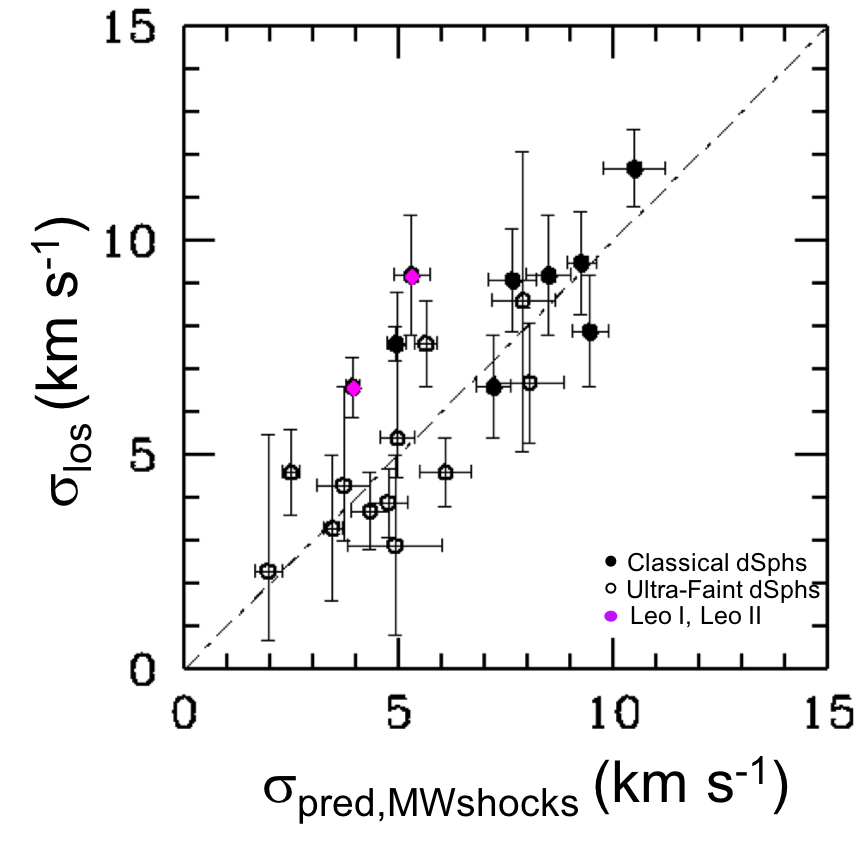}
\end{center}
\caption{\label{fig2} {\it Left:} Tidal shocks (or DM) acceleration (in km $s^{-2}$) based on dSph kinematics ($\Delta \sigma_{los}^2 \times r_{\rm half}^{-1}$) versus MW distance (in kpc). Data ($\sigma_{\rm los}$, $L_{\rm V}$, $r_{\rm half}$) are coming from \cite{Hammer2019,Hammer2020} and the amplitude of $\Delta \sigma_{los}^2 \times r_{\rm half}^{-1}$ assumes $f_{\rm MWshocks} \approx 0.25$. The later value is also supported when comparing tidal shocks caused by a MW potential based on \cite{Bovy2015} or \cite{Eilers2019} mass distribution. {\it Right:} Correlation between the observed $\sigma_{\rm los}$ and that predicted from Eq.~\ref{shock} (see also footnote 1), i.e.,  $\sigma_{\rm pred,MWshocks}$ = $\sqrt{\Delta\sigma_{\rm los}^2 + \sigma_{\rm stars}^2}$, assuming the MW mass model from \cite{Eilers2019}. }
\end{figure}
If the DM is responsible of the velocity dispersion excess of dSphs, $a_{\rm DM}$ does not depends on the distance to the MW, because it is caused by the gravity of the dSph DM mass, and also because DM is expected to shield the dSph mass content against MW tides (see also predictions from {\it Via Lactea} simulations \cite{Diemand2008} in Figure 8 of \cite{Hammer2019}).
 If tidal shocks are responsible of the velocity dispersion increase (see right panel of Figure~\ref{fig2}), the gravitational attraction $a_{\rm DM}$ attributed to the DM is in fact equal to 2$\sqrt{2} \times g_{\rm MW} \times f_{\rm MWshocks}$ (see Eq.~\ref{shock}), in which $g_{\rm MW}$ induces a strong dependency to the distance to the MW center ($D_{MW}$).
 
Left panel of Figure~\ref{fig2} shows that $a_{\rm DM}$=$\Delta \sigma_{\rm los}^2 \times r_{\rm half}^{-1}$ anti-correlates with the MW distance for most dSphs having a robust measurement of their internal velocity dispersion ($\sigma_{los}$). The result is not dependent on a specific  MW mass model because all quantities are observed.  Correlation strength is $\rho$= 0.76 for 21 dSph galaxies, leading to a probability of an occurrence at random to be as low as P= 3 $10^{-4}$ (see all details in \cite{Hammer2019}).  

\section{Conclusions}
The tidal shock scenario for most MW dSphs appears to be a serious competitor to the usually adopted DM-dominated satellites/subhalos scenario. This is because:
\begin{itemize}
\item Most MW dSphs are found near their pericenters, which exclude long-lived, DM-dominated satellites (sub-haloes) with an associated probability P=2 $10^{-7}$; this cannot be circumvented by adapting the MW total mass towards high values that are not consistent with the MW rotation curve.
\item The anti-correlation between $a_{\rm DM}$= $\Delta \sigma_{\rm los}^2 \times r_{\rm half}^{-1}$ and $D_{MW}$  (left panel of Figure~\ref{fig2}, probability that it is fortuitous, P=3 $10^{-4}$)  is a natural prediction if dSphs are tidally shocked, while it cannot be reproduced if dSphs were DM-dominated.
\item MW tidal shocks bring sufficient kinetic energy to generate the high velocity dispersions ($\sigma_{\rm los}$) of dSphs (see right panel of Figure~\ref{fig2}), and former DM mass determinations might be a misinterpretation of such a physical process on DM-free dSphs.
\item Tidal shocks are less destructive than tidal stripping since they bring energy to the dSph stellar content that is partly in resonance with the MW gravitational potential; important mass losses are expected but without tails, and this may correspond to the stars in the dSphs outskirts discovered in wide field observations of classical dSphs (see \cite{McMonigal2014} and references therein).
\item After the first passage at pericenter, tides become increasingly dominated by stripping and it is why, e.g., Sagittarius, Bootes I and Crater II escape the anti-correlation (see Figure 6 of \cite{Hammer2020}).
\end{itemize}
The MW dSphs are either exceptional or $\Lambda$CDM predictions at the low mass end need to be revisited. Star formation histories of few classical dSphs showing neither young nor intermediate-age stars may appear at odd with a first infall. However, such counter argument deserves to be re-investigated because gas-rich dwarfs having stellar mass equal or lower than that of Sculptor mass may have had their star formation stopped by several mechanisms \cite{Garrison-Kimmel2019}, and the absence of DM would let stellar winds more efficient in removing the gas and stopping the star formation. \\

The tidal shock scenario may also apply to the infall of gas-rich stellar systems into a massive body, galaxy, group or cluster of galaxies. A gas-rich galactic system is first stripped of most of its gas leading to a spherical expansion of the residual stars, which leads to tidal shocks when it is reaching its orbital pericenter.


\bibliographystyle{JHEP} 
\bibliography{references_hammer.bib}

\printindex

\end{document}